\documentstyle[11pt,epsf]{article}

\def\indspace{\hspace*{1.0em} }
\def\appendix{\setcounter{section}{0}
\def\thesection{Appendix \Alph{section}}
\def\theequation{\Alph{section}.\arabic{equation}}}

\newfont{\subsub}{cmr6}

\newcounter{szk}

\begin{document}
\title{Derivation of the distribution from extended Gibrat's law
}
\author{
\footnote{e-mail address: ishikawa@kanazawa-gu.ac.jp} Atushi Ishikawa
\\
Kanazawa Gakuin University, Kanazawa 920-1392, Japan
}
\date{}
\maketitle

\begin{abstract}
\indent
Employing profits data of Japanese companies in 2002 and 2003,
we identify the non-Gibrat's law which holds in the middle profits region.
From the law of detailed balance in all regions, Gibrat's law in the high region 
and the non-Gibrat's law in the middle region,
we kinematically derive the profits distribution function in the high and middle range
uniformly.
The distribution function  accurately
fits with empirical data 
without any fitting parameter.
\end{abstract}
\begin{flushleft}
\end{flushleft}

\vspace{1cm}
\section{Introduction}
\label{sec-Introduction}
\indspace
The distribution of wealth, income or company size
is one of important issues not only in economics but also in econophysics
\cite{MS}.
In these distributions,
a cumulative number $N(> x)$ obeys a power-law
for $x$ which is larger than a certain 
threshold $x_0$:
\begin{eqnarray}
    N(> x) \propto x^{-\mu}~~~~{\rm for }~~~~x > x_0~.
    \label{Pareto}
\end{eqnarray}
This power-law and the exponent $\mu$ are called 
Pareto's law and Pareto index, respectively \cite{Pareto} \footnote{
Recently Pareto's law is checked with high accuracy (See \cite{ASNOTT, Yakovenko} for instance).
}.
Here $x$ is wealth, income, 
profits, assets, sales, the number of employees and etc.

The study of power-law distributions in the high range is quite significant.
Because 
a large part of the total wealth, income or profits
is occupied by persons or companies in the high region,
although the number of them is a few percent.
They have the possibility to influence economics.
Power-law distributions are observed
in fractal systems which have self-similarity.
This means that there is self-similarity in economic systems.
The power-law distribution in the high region
is well investigated by using various models in econophysics.

Recently,
Fujiwara et al. \cite{FSAKA} find that
Pareto's law (and the reflection law)
can be derived kinematically
form the law of detailed balance and Gibrat's law \cite{Gibrat}
which are observed in high region $x > x_0$.
In the proof, they assume no model and only use
these two underlying laws in empirical data.

The detailed balance is time-reversal symmetry:
\begin{eqnarray}
    P_{1 2}(x_1, x_2) = P_{1 2}(x_2, x_1)~.
    \label{Detailed balance}
\end{eqnarray}
Here $x_1$ and $x_2$ are  two successive incomes, 
profits, assets, sales, etc, and
$P_{1 2}(x_1, x_2)$ is a joint probability distribution function (pdf).
Gibrat's law states that
the conditional probability distribution of growth rate $Q(R|x_1)$ 
is independent of the initial value $x_1$: 
\begin{eqnarray}
    Q(R|x_1) = Q(R)~.
    \label{Gibrat}
\end{eqnarray}
Here growth rate $R$ is defined as the ratio $R = x_2/x_1$ and
$Q(R|x_1)$ is defined by using the pdf $P(x_1)$
and the joint pdf $P_{1 R}(x_1, R)$ as
\begin{eqnarray}
    Q(R|x_1) = \frac{P_{1 R}(x_1, R)}{P(x_1)}~.
    \label{conditional}
\end{eqnarray}

\begin{figure}[hbp]
 \centerline{\epsfxsize=0.8\textwidth\epsfbox{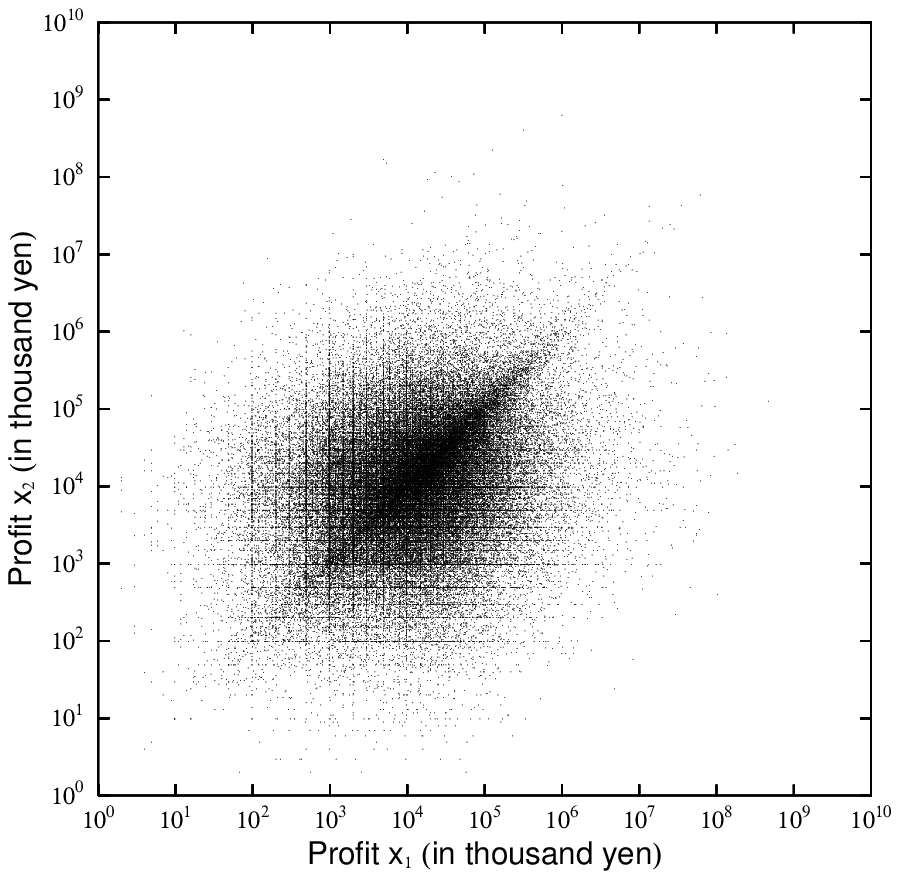}}
 \caption{The scatter plot of all companies, the profits of which in 2002 ($x_1$) and 2003 ($x_2$)
 exceeded $0$, $x_1 > 0$ and $x_2 > 0$.
 The number of the companies is ``132,499''.}
 \label{Profit2002vsProfit2003}
\end{figure}

In Ref. \cite{Ishikawa2},
by using profits data of Japanese companies in 2002 ($x_1$) and 2003 ($x_2$),
it is confirmed that
the law of detailed balance (\ref{Detailed balance}) 
holds 
in all regions $x_1 > 0$ and $x_2 > 0$
(Fig.~\ref{Profit2002vsProfit2003}).
We also find that
Gibrat's law still holds in the extended region
$x_1 > x_0$ and $x_2 > 0$.
From these observations,
we have shown that Pareto index
is also induced from the growth rate distribution empirically.

\begin{figure}[hbp]
 \centerline{\epsfxsize=0.8\textwidth\epsfbox{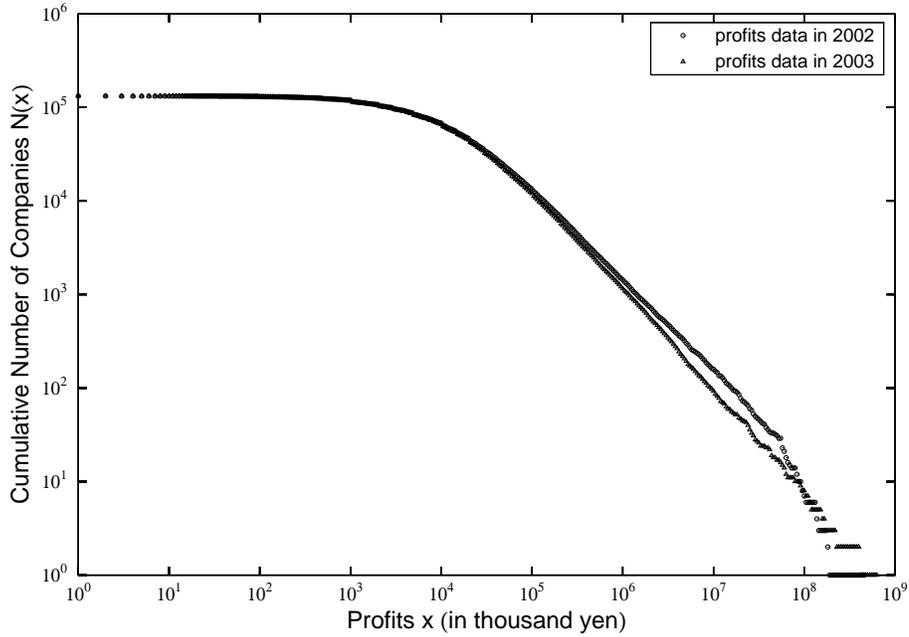}}
 \caption{Cumulative number distributions $N(x_1)$ and $N(x_2)$ for companies, the
 profits of which in 2002 ($x_1$) and 2003 ($x_2$)
 exceeded $0$, $x_1 > 0$ and $x_2 > 0$.
 The number of the companies is ``132,499''.}
 \label{ProfitDistribution}
\end{figure}

On the other hand,
it is also well known that
the power-law is not observed below the threshold $x_0$ \cite{Gibrat, Badger}.
For instance, we show the profits distributions
of Japanese companies in 2002 and 2003 (Fig.~\ref{ProfitDistribution}).
We find that Pareto's law holds in the high profits region
and it fails in the middle one.
The study of distributions in the middle region
is as important as the study of power-law those.
Because
a large number of persons or companies
is included in the middle region.
Furthermore,
it is interesting to study the breaking of fractal.

In order to obtain the distribution in the middle region,
we examine data below the threshold $x_0$.
It is reported that 
Gibrat's law is valid only in the high region (\cite{OTT} for instance).
The recent works about the breakdown of Gibrat's law 
is done by Stanley's group \cite{Stanley1}.
Aoyama et al. also report that Gibrat's law does not hold
in the middle region by using data of Japanese companies
\cite{Aoyama}.

In the analysis of Gibrat's law in Ref.~\cite{Ishikawa2},
we concentrate our attention to the region $x_1 > x_0$ and $x_2 > 0$.
In this paper, 
we examine the growth rate distribution
in all regions $x_1 > 0$ and $x_2 > 0$
and identify the law which is the extension of Gibrat's law \footnote{
In this paper, we call the law in the region
$x_1 < x_0$ and $x_2 > 0$ non-Gibrat's law.
}.
We employ profits data of Japanese companies in 2002 and 2003
which are available on 
the database ``CD Eyes'' published by Tokyo Shoko Research, Ltd.
\cite{TSR}.

By using the extended Gibrat's law,
we derive the distribution function in the high and middle profits region
under the law of detailed balance.
It
explains empirical data 
with high accuracy.
Notice that the distribution function has no fitting parameter.
The parameters of the function are already decided in the extended Gibrat's law.
\section{Gibrat's law and non-Gibrat's law}
\label{sec-Gibrat's law and non-Gibrat's law}
\begin{figure}[htb]
 \centerline{\epsfxsize=0.8\textwidth\epsfbox{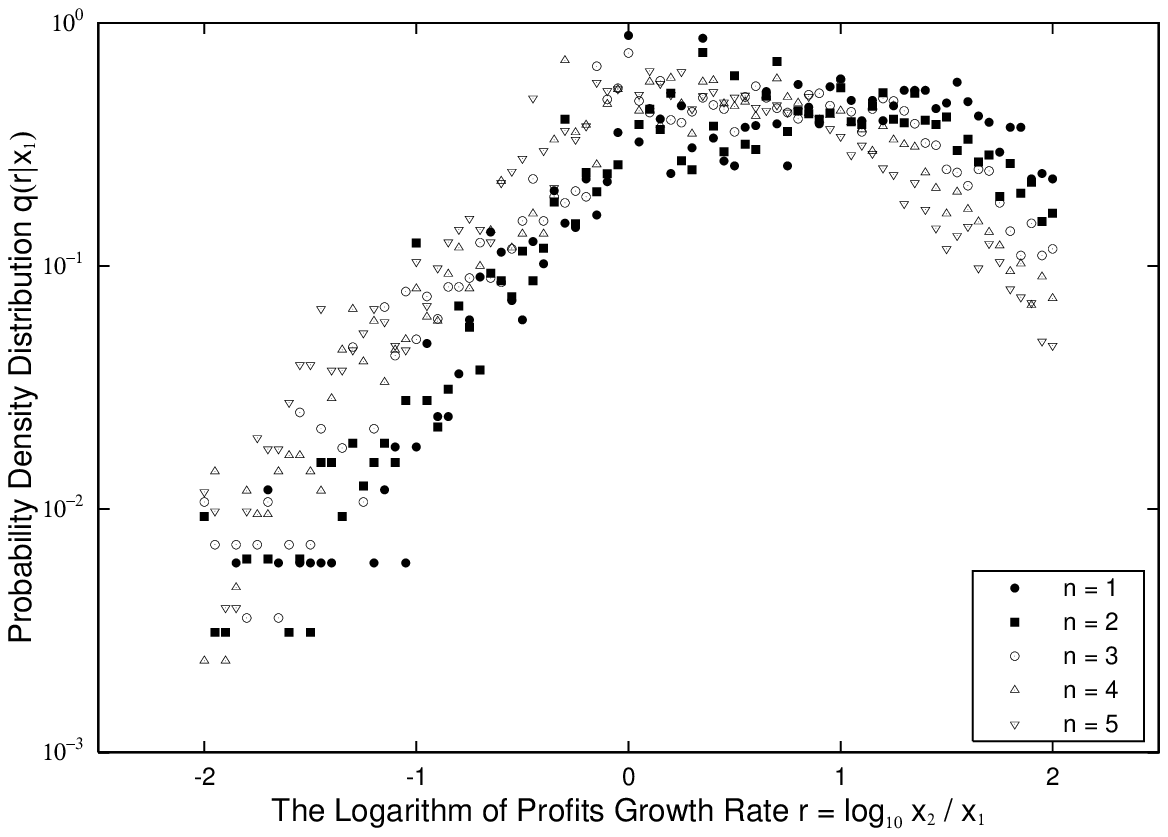}}
 \caption{The probability density distribution $q(r|x_1)$ of the log profits growth rate
 $r = \log_{10} x_2/x_1$ from 2002 to 2003.
 The data points are classified into five 
 bins of the initial profits with equal magnitude in logarithmic scale,
 $x_1 \in 4 \times [10^{2+0.2(n-1)},10^{2+0.2n}]~(n=1, 2, \cdots, 5)$ thousand yen.
 The number of companies in this regime is ``35,513''.}
 \label{ProfitGrowthRateL}
\end{figure}

\begin{figure}[htb]
 \centerline{\epsfxsize=0.8\textwidth\epsfbox{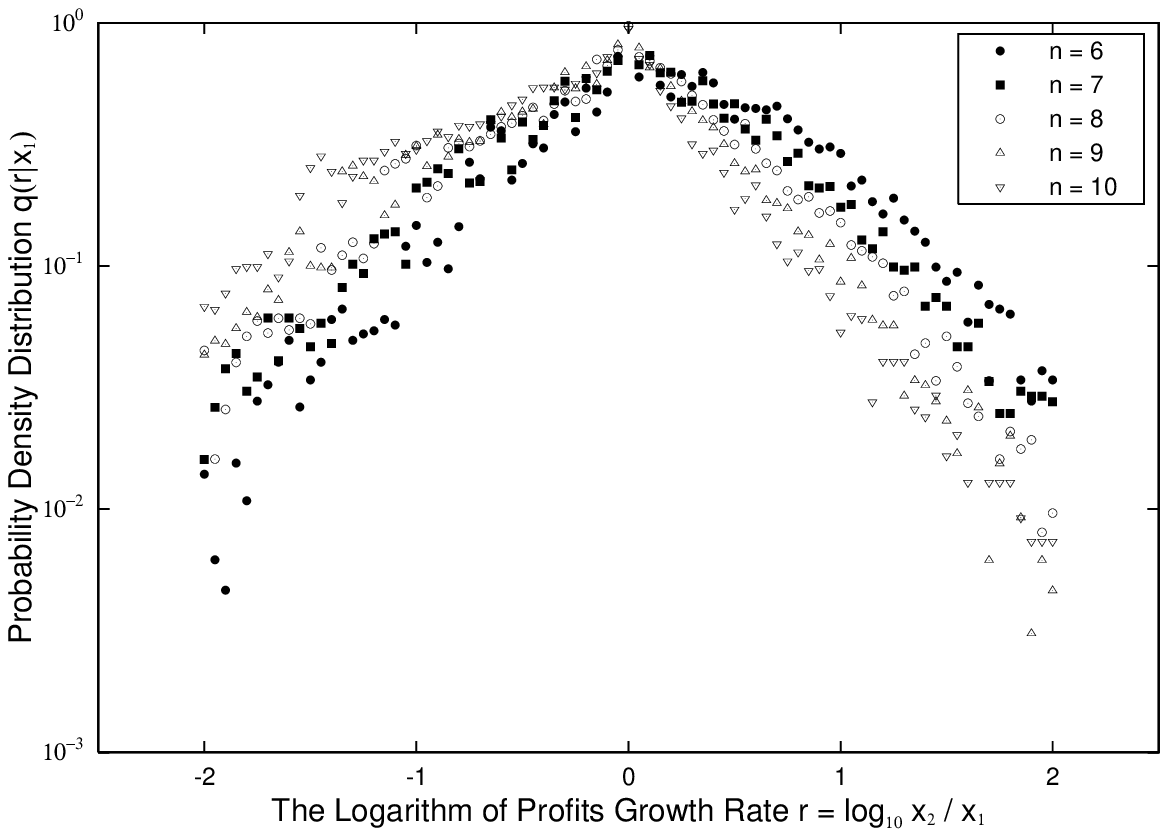}}
 \caption{The probability density distribution $q(r|x_1)$ of the log profits growth rate
 $r= \log_{10} x_2/x_1$ from 2002 to 2003.
 The data points are also classified into five 
 bins of the initial profits with equal magnitude in logarithmic scale,
 $x_1 \in 4 \times [10^{2+0.2(n-1)},10^{2+0.2n}]~(n=6, 7, \cdots, 10)$ thousand yen.
 The number of companies in this regime is ``64,205''.}
 \label{ProfitGrowthRateM}
\end{figure}

\begin{figure}[htb]
 \centerline{\epsfxsize=0.8\textwidth\epsfbox{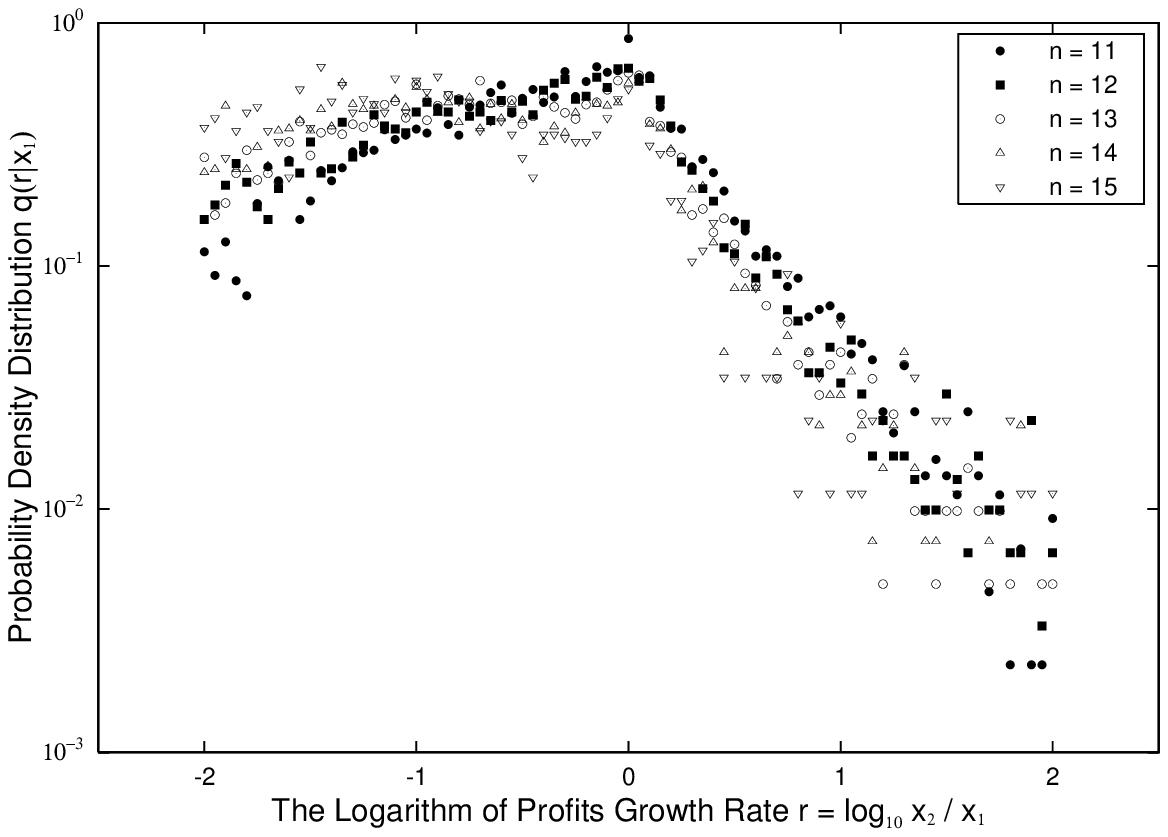}}
 \caption{The probability density distribution $q(r|x_1)$ of the log profits growth rate
 $r= \log_{10} x_2/x_1$ from 2002 to 2003.
 The data points are also classified into five 
 bins of the initial profits with equal magnitude in logarithmic scale,
 $x_1 \in 4 \times [10^{2+0.2(n-1)},10^{2+0.2n}]~(n=11, 12, \cdots, 15)$ thousand yen.
 The number of companies in this regime is ``25,189''.}
 \label{ProfitGrowthRateH}
\end{figure}

\indspace
In this section, we reconfirm Gibrat's law in the high profits region
and identify non-Gibrat's law in the middle one.
We divide the range of $x_1$ into logarithmically equal bins as
$x_1 \in 4 \times [10^{2+0.2(n-1)},10^{2+0.2n}]$ thousand yen
with $n=1, 2, \cdots, 15$ \footnote{
In Ref.~\cite{Ishikawa2}, we only consider the case 
for $n=11, 12, \cdots, 15$.
}.
In Fig.~\ref{ProfitGrowthRateL}, \ref{ProfitGrowthRateM} and
\ref{ProfitGrowthRateH},
the probability densities for $r$ 
are expressed in the case of $n=1, \cdots, 5$,
$n=6, \cdots, 10$ 
and $n=11, \cdots, 15$,
respectively.
The number of the companies in Fig.~\ref{ProfitGrowthRateL}, 
\ref{ProfitGrowthRateM} and \ref{ProfitGrowthRateH}
is ``$35,513$", ``$64,205$" and ``$25,189$", respectively.
Here we use the log profits growth rate $r=\log_{10} R$.
The probability density for $r$ defined by $q(r|x_1)$ is
related to that for $R$ by
\begin{eqnarray}
    \log_{10}Q(R|x_1)+r+\log_{10}(\ln 10)=\log_{10}q(r|x_1)~.
\label{Qandq}
\end{eqnarray}

From Fig.~\ref{ProfitGrowthRateL}, \ref{ProfitGrowthRateM} and
\ref{ProfitGrowthRateH},
we express the relation between $r$ and $q(r|x_1)$
as follows:
\begin{eqnarray}
    \log_{10}q(r|x_1)&=&c(x_1)-t_{+}(x_1)~r~~~~~{\rm for}~~r > 0~,
    \label{approximation1}\\
    \log_{10}q(r|x_1)&=&c(x_1)+t_{-}(x_1)~r~~~~~{\rm for}~~r < 0~.
    \label{approximation2}
\end{eqnarray}
By the use of Eq.~(\ref{Qandq}),
these relations are rewritten in terms of $R$ as
\begin{eqnarray}
    Q(R|x_1)&=&d(x_1)~R^{-t_{+}(x_1)-1}~~~~~{\rm for}~~R > 1~,
    \label{Rline1}\\
    Q(R|x_1)&=&d(x_1)~R^{+t_{-}(x_1)-1}~~~~~{\rm for}~~R < 1~,
    \label{Rline2}
\end{eqnarray}
with $d(x_1)=10^{c(x_1)}/{\ln 10}$~.

\begin{figure}[htb]
 \centerline{\epsfxsize=0.8\textwidth\epsfbox{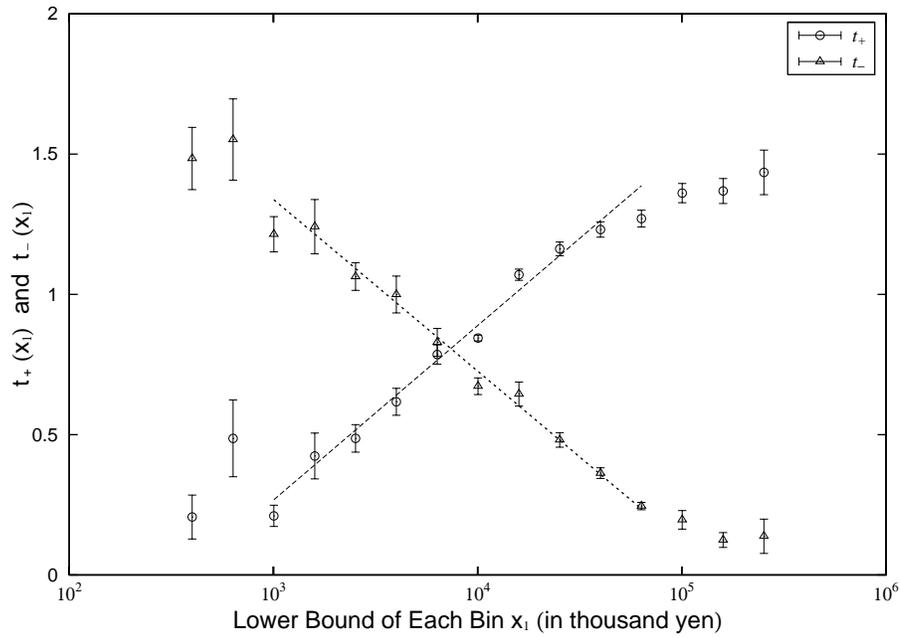}}
 \caption{
 The relation between the lower bound of each bin $x_1$ and $t_{\pm}(x_1)$.
 From the left, each data point represents $n=1, 2, \cdots, 15$.
 }
 \label{X1vsT}
\end{figure}

By applying the expressions (\ref{approximation1}) and (\ref{approximation2}) 
to data in 
Fig.~\ref{ProfitGrowthRateL}, \ref{ProfitGrowthRateM} and
\ref{ProfitGrowthRateH},
the relation between $x_1$ and $t_{\pm}(x_1)$ 
is obtained (Fig.~\ref{X1vsT}).
In Fig.~\ref{X1vsT}, 
$t_{\pm}$ hardly responds to 
$x_1$ for $n=13, 14, 15$.
This means that Gibrat's law holds in the high profits region.
On the other hand,
$t_{+}$ ($t_{-}$) increases (decreases) linearly with $\log_{10} x_1$
for $n=4, 5, \cdots, 11$.
In the middle profits region, 
not Gibrat's law but the other law (non-Gibrat's law) holds as follows \footnote{
In section \ref{sec-Another non-Gibrat's law},
we examine another type of non-Gibrat's law.
}:
\begin{eqnarray}
    t_{+}(x_1)&=&t_{+}(x_0)-\alpha_{+}~\ln \frac{x_1}{x_0}~,
    \label{t+}\\
    t_{-}(x_1)&=&t_{-}(x_0)+\alpha_{-}~\ln \frac{x_1}{x_0}~.
    \label{t-}
\end{eqnarray}
In Fig.~\ref{X1vsT}, $\alpha_{+}$ and $\alpha_{-}$ are estimated
as
\begin{eqnarray}
    \alpha_{+} \sim \alpha_{-} &\sim& 0~~~~~~~~~{\rm for}~~x_1 > x_0~,
    \label{alphaH}\\
    \alpha_{+} \sim \alpha_{-} &\sim& 0.27~~~~~{\rm for}~~x_{{\rm min}} < x_1 < x_0~,
    \label{alphaM}
\end{eqnarray}
where $x_0 = 4 \times 10^{2+0.2(13-1)} \sim 10^5$ thousand yen ($= 100$ million yen)
and $x_{{\rm min}} = 4 \times 10^{2+0.2(4-1)} \sim 1,600$ thousand yen.
In this paper we call the combination of Gibrat's law 
((\ref{t+}), (\ref{t-}) and (\ref{alphaH}))
and non-Gibrat's law
((\ref{t+}), (\ref{t-}) and (\ref{alphaM}))
extended Gibrat's law.

\section{Profits distribution in the high  and middle region}
\label{sec-Profits distribution in the high  and middle region}
\indspace
In Refs.~\cite{Ishikawa2},
Pareto's law (\ref{Pareto}) and the Pareto index $\mu$ can be derived
from the detailed balance (\ref{Detailed balance}) and
Gibrat's law (\ref{Gibrat})
in the high profits region.
In this section, we derive profits distribution not only
in the high profits region but also in the middle one
by using the detailed balance and the extended Gibrat's law.

Due to the relation of
$P_{1 2}(x_1, x_2)dx_1 dx_2 = P_{1 R}(x_1, R)dx_1 dR$
under the change of variables from $(x_1, x_2)$ to $(x_1, R)$,
these two joint pdfs are related to each other,
\begin{eqnarray}
    P_{1 R}(x_1, R) = x_1 P_{1 2}(x_1, x_2).
\end{eqnarray}
By the use of this relation, the detailed balance (\ref{Detailed balance})
is rewritten in terms of $P_{1 R}(x_1, R)$ as follows:
\begin{eqnarray}
    P_{1 R}(x_1, R) = R^{-1} P_{1 R}(x_2, R^{-1}).
\end{eqnarray}
Substituting the joint pdf $P_{1 R}(x_1, R)$ for the conditional probability $Q(R|x_1)$
defined in Eq.~(\ref{conditional}),
the detailed balance is expressed as
\begin{eqnarray}
    \frac{P(x_1)}{P(x_2)} = \frac{1}{R} \frac{Q(R^{-1}|x_2)}{Q(R|x_1)}~.
\end{eqnarray}

In the preceding section, the conditional probability $Q(R|x_1)$
is identified as (\ref{Rline1}) or (\ref{Rline2}).
Under the change of variables $x_1 \leftrightarrow x_2$,
the conditional probability $Q(R^{-1}|x_2)$ is expressed as
\begin{eqnarray}
    Q(R^{-1}|x_2)&=&d(x_2)~R^{-t_{+}(x_2)-1}~~~~~{\rm for}~~R < 1~,\\
    Q(R^{-1}|x_2)&=&d(x_2)~R^{+t_{-}(x_2)-1}~~~~~{\rm for}~~R > 1~.
\label{Rline}
\end{eqnarray}

From the detailed balance 
and the extended Gibrat's law,
one finds the following:
\begin{eqnarray}
    \frac{P(x_1)}{P(x_2)} &=& R^{+t_{+}(x_1)-t_{-}(x_2)+1}\\
    &=& R^{+t_{+}(x_0)-t_{-}(x_0)+1+\alpha_{+} \ln \frac{x_1}{x_0}
            +\alpha_{-} \ln \frac{x_2}{x_0}}
    \label{DE0}
\end{eqnarray}
for $R>1$.
Here we assume that the $x$ dependence of $d(x)$ 
is negligible in the derivation,
the validity of which should be checked against the results.
By expanding Eq.~(\ref{DE0}) around $R=1$, the
following differential equation is obtained 
\begin{eqnarray}
    \left[+t_{+}(x_0)-t_{-}(x_0)
        +1+\left(\alpha_{+}+\alpha_{-}\right)\ln \frac{x}{x_0}\right] P(x) 
        + x~ P'(x) = 0,
\end{eqnarray}
where $x$ denotes $x_1$.
The same differential equation is obtained for $R<1$.
The solution is given by
\begin{eqnarray}
    P(x) = C x^{-\left(\mu+1\right)}~e^{-\alpha \ln^2 \frac{x}{x_0}}
    \label{HandM}
\end{eqnarray}
with $\alpha = \left(\alpha_{+}+\alpha_{-}\right)/2$.
Here we use the relation $+t_{+}(x_0)-t_{-}(x_0) = \mu$
which is confirmed in Ref.~\cite{Ishikawa2}.
\section{Data fitting}
\label{sec-Data Fitting}
\indspace
In the previous section, we derive the profits distribution function (\ref{HandM})
in the high and middle profits region
from the detailed balance and the extended Gibrat's law.
In this section,
we directly examine whether it fits with profits distribution data.

In order to average the scattering of data points,
we employ the cumulative number of companies (Fig.~\ref{ProfitDistribution}).
For $x_1 > x_0$,
by using Eqs.~(\ref{alphaH}) and (\ref{HandM})
the cumulative distribution in the high profits region
is expressed as
\begin{eqnarray}
    N_{\rm H}(>x) &=& N(>x_0) \int^{\infty}_{x}~dt~P(t)\\
                  &=& N(>x_0) \left(\frac{x}{x_0}\right)^{-\mu}.
\label{CDH}
\end{eqnarray}

On the other hand, by the use of Eqs.~(\ref{alphaM}) and (\ref{HandM})
the cumulative distribution in the middle profits region
$x_{{\rm min}} < x_1 < x_0$ is given by
\begin{eqnarray}
    N_{\rm M}(>x) &=& \left\{N(>x_{\rm min})-N(>x_0)\right\}
                         \int^{x_0}_{x}~dt~P(t) + N(>x_0)\\
                  &=& \left\{N(>x_{\rm min})-N(>x_0)\right\}
                        \frac{{\rm Erf}\left(\frac{1}{2 \sqrt{\alpha}}\right)
                        -{\rm Erf}\left(\frac{\mu+2 \alpha \ln \frac{x}{x_0}}
                                {2 \sqrt{\alpha}}\right)}
                        {{\rm Erf}\left(\frac{1}{2 \sqrt{\alpha}}\right)
                        -{\rm Erf}\left(\frac{\mu+2 \alpha \ln \frac{x_{\rm min}}{x_0}}
                                {2 \sqrt{\alpha}}\right)} \nonumber\\
                  && +~N(>x_0)~.
    \label{CDM}
\end{eqnarray}
Here ${\rm Erf}(x)$ is error function defined by 
${\rm Erf}(x) = 2 \int^{x}_{0} e^{-t^2} dt/\sqrt{\pi}$.

\begin{figure}[htb]
 \centerline{\epsfxsize=0.8\textwidth\epsfbox{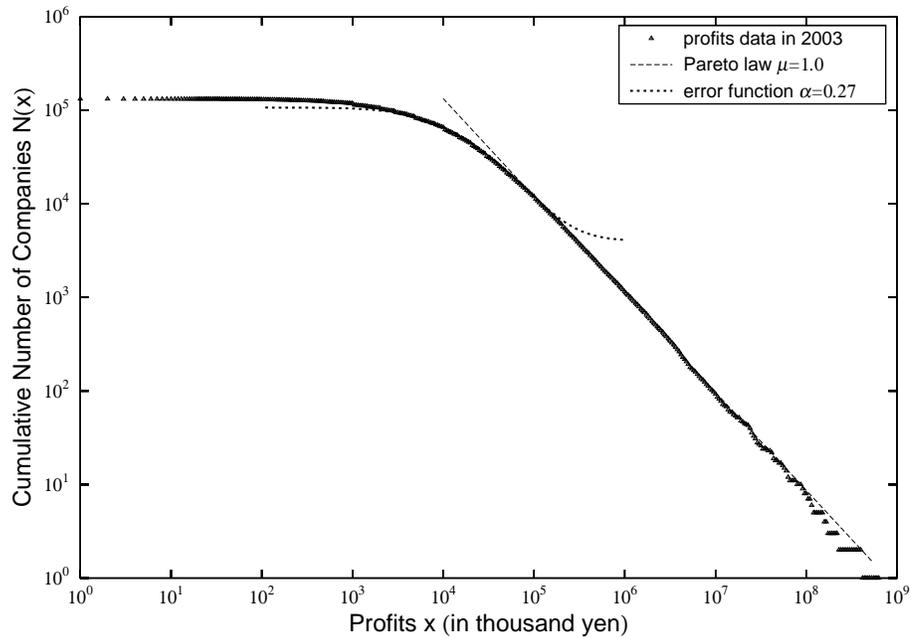}}
 \caption{
 Cumulative number distributions $N(x_2)$ for companies, the
 profits of which in 2002 ($x_1$) and 2003 ($x_2$)
 exceeded $0$, $x_1 > 0$ and $x_2 > 0$.
 The cumulative distribution function derived from the detailed balance
 and the extended Gibrat's law accurately fits with the data.
  Indices $\mu$, $\alpha$ and the bounds $x_0$, $x_{\rm min}$ is already
  given in the extended Gibrat's law.
 }
 \label{ProfitDistributionFit}
\end{figure}

In Fig.~\ref{ProfitDistributionFit},
the distribution functions (\ref{CDH}) in the high profits region ($x_1 > x_0$)
and (\ref{CDM}) in the middle one ($x_{{\rm min}} < x_1 < x_0$)
explain empirical data in 2003 with high accuracy.
This guarantees the validity of the assumption in the previous section.
Notice that
there is no ambiguity in parameter fitting,
because indices $\mu$, $\alpha$ and 
the bounds $x_0$, $x_{\rm min}$ is already
given in the extended Gibrat's law \footnote{
Pareto's law with $\mu=1$ is especially called Zipf's law \cite{Zipf}.
}.

\section{Another non-Gibrat's law}
\label{sec-Another non-Gibrat's law}
\indspace
In section \ref{sec-Gibrat's law and non-Gibrat's law},
we present the non-Gibrat's law (\ref{t+}), (\ref{t-}) and (\ref{alphaM})
as a linear approximation of $t_{\pm}(x_1)$ in Fig.~\ref{X1vsT}.
In this section,
we examine another linear approximation in Fig.~\ref{X1vsT2},
the vertical axis of which is the logarithm.

In Fig.~\ref{X1vsT2},
$\log_{10} t_{+}$ ($\log_{10} t_{-}$) increases (decreases) linearly 
with $\log_{10} x_1$ for $n=4, 5, \cdots, 11$.
This relation is expressed as
\begin{eqnarray}
    t_{+}(x_1)&=&t_{+}(x_0)~\left( \frac{x_1}{x_0} \right)^{+\beta_{+}},
    \label{t+another}\\
    t_{-}(x_1)&=&t_{-}(x_0)~\left( \frac{x_1}{x_0} \right)^{-\beta_{-}}~,
    \label{t-another}
\end{eqnarray}
where
\begin{eqnarray}
    t_{+}(x_0) - t_{-}(x_0) &=& \mu~,\\
    \beta_{+} \sim 0.35~,~~\beta_{-} &\sim& 0.36~~~~~{\rm for}~~x_{{\rm min}} < x_1 < x_0~,\\
    \beta_{+} \sim \beta_{-} &\sim& 0~~~~~~~~~{\rm for}~~x_1 > x_0~.
    \label{}
\end{eqnarray}
Here $x_0$ and $x_{{\rm min}}$ are same values in section 
\ref{sec-Gibrat's law and non-Gibrat's law}.
From
the detailed balance 
and this extended Gibrat's law, 
one finds
\begin{eqnarray}
    \frac{P(x_1)}{P(x_2)} &=& R^{+t_{+}(x_0)~\left( \frac{x_1}{x_0} \right)^{+\beta_{+}}
                -t_{-}(x_0)~\left( \frac{x_2}{x_0} \right)^{-\beta_{-}}+1}
    \label{DandG}
\end{eqnarray}
for $R > 1$.

By expanding this equation around $R=1$, the
following differential equation is obtained 
\begin{eqnarray}
    \left[+t_{+}(x_0)~\left( \frac{x}{x_0} \right)^{+\beta_{+}}
                -t_{-}(x_0)~\left( \frac{x}{x_0} \right)^{-\beta_{-}}+1 \right] P(x) 
        + x~ P'(x) = 0~,
    \label{DE}        
\end{eqnarray}
where $x$ denotes $x_1$.
The same differential equation is obtained for $R<1$.
In order to take $\beta_{\pm} \to 0$ limit into account,
we rewrite this as follows:
\begin{eqnarray}
    \left[+t_{+}(x_0)\left\{\left( \frac{x}{x_0} \right)^{+\beta_{+}}-1\right\}
                -t_{-}(x_0)\left\{\left( \frac{x}{x_0} \right)^{-\beta_{-}}-1\right\}
                +\mu+1 \right] P(x) 
        + x~ P'(x) = 0~.
\end{eqnarray}
The solution is given by
\begin{eqnarray}
    P(x) = C x^{-\left(\mu+1\right)}~e^{-\beta \ln^2 \frac{x}{x_0}}
            +O({\beta_+}^2)+O({\beta_-}^2)
    \label{HandManother}
\end{eqnarray}
with $\beta = \left(\beta_{+}t_{+}(x_0)+\beta_{-}t_{-}(x_0)\right)/2$.
Here we use the expansion 
$x^{\beta-1}/(x_0)^{\beta} - x^{-1} = \beta (\ln x/x_0)/x + O(\beta^2)$.

If we neglect $O({\beta_{\pm}}^2)$ terms in Eq.~(\ref{HandManother}),
the profits distribution function (\ref{HandManother}) can be identified with
(\ref{HandM}),
because $\beta \sim \alpha$ numerically.
In other words,
there is no essential difference between two expressions (\ref{t+}), (\ref{t-})
and (\ref{t+another}), (\ref{t-another})
in the middle region.
\section{Conclusion}
\label{sec-Conclusion}
\indspace
In this paper,
we have kinematically derived the profits distribution function
in the high and middle region
from the law of detailed balance and the extended Gibrat's law
by employing profits data of Japanese companies
in 2002 and 2003.

\begin{figure}[htb]
 \centerline{\epsfxsize=0.8\textwidth\epsfbox{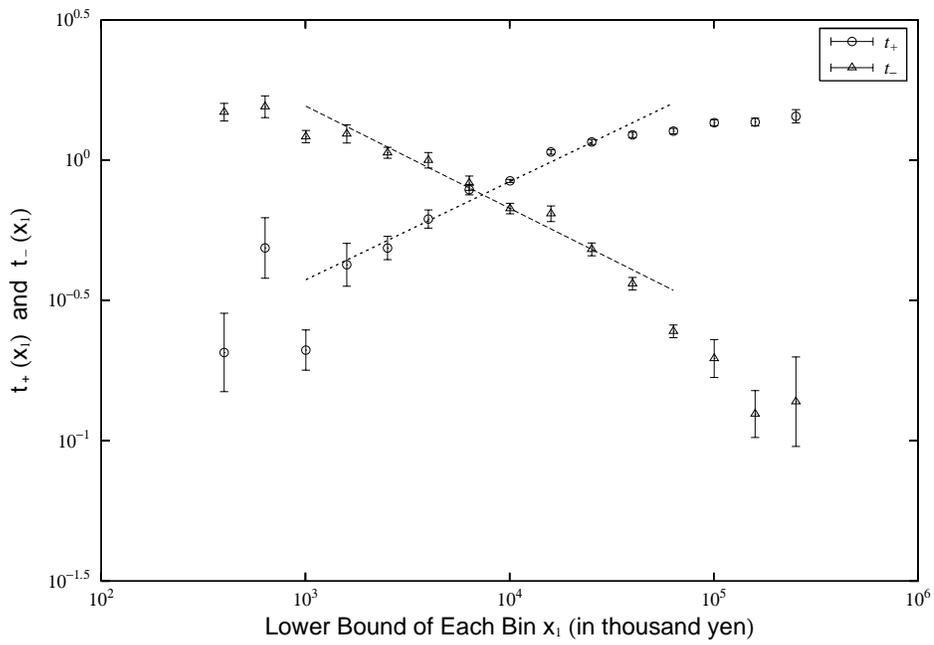}}
 \caption{
    The log-log plot of the relation 
    between the lower bound of each bin $x_1$ and $t_{\pm}(x_1)$.
    From the left, each data point represents $n=1, 2, \cdots, 15$.
 }
 \label{X1vsT2}
\end{figure}

Firstly, we have reconfirmed Gibrat's law in the high profits region
($x > x_0$). The value of $x_0$ is estimated to be about 
$100$ million yen in Fig.~\ref{X1vsT} or \ref{X1vsT2}.
At the same time,
we have found that Gibrat's law fails and another law holds
in the middle profits region
($x_{{\rm min}} < x < x_0$).
We have identified the non-Gibrat's law
and the value of $x_{\rm min}$ is estimated to be about 
$1,600$ thousand yen
in Fig.~\ref{X1vsT} or \ref{X1vsT2}.
We have called the combination of Gibrat's law and the non-Gibrat's law the
extended Gibrat's law.

Secondly, we have derived not only Pareto's law in the high profits region
but also the distribution in the middle one
from the detailed balance and the extended Gibrat's law.
The derivation has been described uniformly
in terms of the extended Gibrat's law.

The profits distribution in the middle region
is very similar to log-normal one:
\begin{eqnarray}
    P(x) = C x^{-1}~\exp \left[-\frac{1}{2 \sigma^2} \ln^2 \frac{x}{\bar{x}}\right]~,
    \label{log-normal}
\end{eqnarray}
where $\bar{x}$ is mean value and $\sigma^2$ is variance.
The difference between two distributions (\ref{HandM}) and (\ref{log-normal})
is only the power of $x$.
It brings a translation along the horizontal axis
in the log-log plot.
In this sense, 
the distribution in the middle profits region obtained in this paper
is essentially equivalent to the log-normal one. 
Notice that it has no fitting parameter
that the log-normal distribution, prepared only for data fitting, has.
Indies $\mu$, $\alpha$ and 
the bounds $x_0$, $x_{\rm min}$ is already
given in the extended Gibrat's law.

In the derivation of the profits distribution,
we have used the detailed balance and the extended Gibrat's law.
The detailed balance is observed in a relatively stable period
in economy \cite{FSAKA}.
The extended Gibrat's law is interpreted 
as follows.
Companies, classified in small-scale profits category,
have more (less) possibilities of increasing (decreasing) their profits
than companies classified in large-scale one.
In other words,
it is probably difficult that companies gain large-scale profits
in two successive years.
This is the non-Gibrat's law and 
it leads the distribution in the middle profits region.

This phenomenon is not observed above the threshold 
$x_0$.
For $x > x_0$,
companies, classified in small-scale profits category,
have same possibilities of increasing (decreasing) their profits
with companies classified in large-scale one.
This is the Gibrat's law and 
it leads the power distribution in the high profits region.

In this paper,
we cannot mention the distribution in the low profits region $x < x_{\rm min}$,
because no law is observed in the region (Fig.~\ref{X1vsT} or \ref{X1vsT2}).
This may be caused by insufficient data in the low region.
Lastly, we speculate possible distributions in the region.
If we do not
take $\beta_{\pm} \to 0$ limit,
we should solve the differential equation (\ref{DE}).
The solution is given by
\begin{eqnarray}
    P(x) = C x^{-1}~\exp\left[
                -\frac{t_{+}(x_0)}{\beta_{+}}\left( \frac{x}{x_0} \right)^{+\beta_{+}}
                -\frac{t_{-}(x_0)}{\beta_{-}}\left( \frac{x}{x_0} \right)^{-\beta_{-}}
            \right]~.
    \label{ProfitDistributionL}
\end{eqnarray}
For the case $\beta_{+} \sim 1$ and $\beta_{-} \sim 0$,
the distribution (\ref{ProfitDistributionL}) takes the exponential form 
in Refs.~\cite{Yakovenko, NS}.
On the other hand,
for the case $\beta_{+} \sim 0$ and $\beta_{-} \sim 1$,
the distribution (\ref{ProfitDistributionL}) takes Weibull form in Ref.~\cite{AIST}.
We can decide the distribution in the low profits region
if sufficient data in the region are provided.

We have showed that a distribution function is decided by
underlying kinematics.
For profits data we used, the distribution is power in the high region
and log-normal type in the middle one.
This does not claim that
all the distributions in the middle region are log-normal types.
For instance, the personal income distribution may
take a different form, if the extended Gibrat's law changes.
Even in the case,
the other extended Gibrat's law will decide the other distribution
by the use of the method in this paper.

\section*{Acknowledgments}
\indent

The author is 
grateful to 
Professor~H. Aoyama for useful discussions
about his lecture.





\end{document}